\documentclass[aps,prl,showpacs,showkeys,amsmath,amssymb,
twocolumn,floatfix,superscriptaddress,nolongbibliography]{revtex4-2} 
\usepackage{graphicx}
\usepackage{dcolumn}
\usepackage{bm}
\usepackage{xcolor}
\usepackage{tabularx} 
\usepackage[normalem]{ulem} 
\usepackage[mathlines]{lineno}

\newcommand{\mmdmee}{\Delta {m}^{2}_{ee}} 
\newcommand{\dmee}{$\mmdmee$}
\newcommand{\mmdmtt}{\Delta {m}^{2}_{32}} 
\newcommand{\dmtt}{$\mmdmtt$}
\newcommand{\mmthet}{{\rm sin}^{2}2\theta_{13}} 
\newcommand{\thet}{$\mmthet$}

\newcommand{\nuebar}{$\overline{\nu}_{e}$~}

\newcommand{\TOTALIBD}{$5.55\times10^6$}
\newcommand{\SINTONETHREE}{0.0851\pm0.0024} 
\newcommand{\DMEE}{(2.519\pm0.060)\times 10^{-3}\ {\rm eV}^2}  
\newcommand{\DMTHREETWONORMAL}{(2.466\pm0.060 )\times 10^{-3}\ {\rm eV}^2} 
\newcommand{\DMTHREETWOINVERTED}{-(2.571\pm0.060)\times 10^{-3}\ {\rm eV}^2} 

\begin{document}


\title{Precision measurement of reactor antineutrino oscillation \\ 
at kilometer-scale baselines by Daya Bay}

\newcommand{\IHEP}{\affiliation{Institute~of~High~Energy~Physics, Beijing}}
\newcommand{\Wisconsin}{\affiliation{University~of~Wisconsin, Madison, Wisconsin 53706}}
\newcommand{\Yale}{\affiliation{Wright~Laboratory and Department~of~Physics, Yale~University, New~Haven, Connecticut 06520}} 
\newcommand{\BNL}{\affiliation{Brookhaven~National~Laboratory, Upton, New York 11973}}
\newcommand{\NTU}{\affiliation{Department of Physics, National~Taiwan~University, Taipei}}
\newcommand{\NUU}{\affiliation{National~United~University, Miao-Li}}
\newcommand{\Dubna}{\affiliation{Joint~Institute~for~Nuclear~Research, Dubna, Moscow~Region}}
\newcommand{\CalTech}{\affiliation{California~Institute~of~Technology, Pasadena, California 91125}}
\newcommand{\CUHK}{\affiliation{Chinese~University~of~Hong~Kong, Hong~Kong}}
\newcommand{\NCTU}{\affiliation{Institute~of~Physics, National~Chiao-Tung~University, Hsinchu}}
\newcommand{\NJU}{\affiliation{Nanjing~University, Nanjing}}
\newcommand{\TsingHua}{\affiliation{Department~of~Engineering~Physics, Tsinghua~University, Beijing}}
\newcommand{\SZU}{\affiliation{Shenzhen~University, Shenzhen}}
\newcommand{\NCEPU}{\affiliation{North~China~Electric~Power~University, Beijing}}
\newcommand{\Siena}{\affiliation{Siena~College, Loudonville, New York  12211}}
\newcommand{\IIT}{\affiliation{Department of Physics, Illinois~Institute~of~Technology, Chicago, Illinois  60616}}
\newcommand{\LBNL}{\affiliation{Lawrence~Berkeley~National~Laboratory, Berkeley, California 94720}}
\newcommand{\UIUC}{\affiliation{Department of Physics, University~of~Illinois~at~Urbana-Champaign, Urbana, Illinois 61801}}
\newcommand{\SJTU}{\affiliation{Department of Physics and Astronomy, Shanghai Jiao Tong University, Shanghai Laboratory for Particle Physics and Cosmology, Shanghai}}
\newcommand{\BNU}{\affiliation{Beijing~Normal~University, Beijing}}
\newcommand{\WM}{\affiliation{College~of~William~and~Mary, Williamsburg, Virginia  23187}}
\newcommand{\Princeton}{\affiliation{Joseph Henry Laboratories, Princeton~University, Princeton, New~Jersey 08544}}
\newcommand{\VirginiaTech}{\affiliation{Center for Neutrino Physics, Virginia~Tech, Blacksburg, Virginia  24061}}
\newcommand{\CIAE}{\affiliation{China~Institute~of~Atomic~Energy, Beijing}}
\newcommand{\SDU}{\affiliation{Shandong~University, Jinan}}
\newcommand{\NanKai}{\affiliation{School of Physics, Nankai~University, Tianjin}}
\newcommand{\UC}{\affiliation{Department of Physics, University~of~Cincinnati, Cincinnati, Ohio 45221}}
\newcommand{\DGUT}{\affiliation{Dongguan~University~of~Technology, Dongguan}}
\newcommand{\XJTU}{\affiliation{Department of Nuclear Science and Technology, School of Energy and Power Engineering, Xi'an Jiaotong University, Xi'an}}
\newcommand{\UCB}{\affiliation{Department of Physics, University~of~California, Berkeley, California  94720}}
\newcommand{\HKU}{\affiliation{Department of Physics, The~University~of~Hong~Kong, Pokfulam, Hong~Kong}}
\newcommand{\Charles}{\affiliation{Charles~University, Faculty~of~Mathematics~and~Physics, Prague}} 
\newcommand{\USTC}{\affiliation{University~of~Science~and~Technology~of~China, Hefei}}
\newcommand{\TempleUniversity}{\affiliation{Department~of~Physics, College~of~Science~and~Technology, Temple~University, Philadelphia, Pennsylvania  19122}}
\newcommand{\CGNPG}{\affiliation{China General Nuclear Power Group, Shenzhen}}
\newcommand{\NUDT}{\affiliation{College of Electronic Science and Engineering, National University of Defense Technology, Changsha}} 
\newcommand{\IowaState}{\affiliation{Iowa~State~University, Ames, Iowa  50011}}
\newcommand{\ZSU}{\affiliation{Sun Yat-Sen (Zhongshan) University, Guangzhou}}
\newcommand{\CQU}{\affiliation{Chongqing University, Chongqing}} 
\newcommand{\BCC}{\altaffiliation[Now at ]{Department of Chemistry and Chemical Technology, Bronx Community College, Bronx, New York  10453}} 

\newcommand{\UCI}{\affiliation{Department of Physics and Astronomy, University of California, Irvine, California 92697}} 
\newcommand{\GXU}{\affiliation{Guangxi University, No.100 Daxue East Road, Nanning}} 
\newcommand{\HKUST}{\affiliation{The Hong Kong University of Science and Technology, Clear Water Bay, Hong Kong}} 
\newcommand{\Rochester}{\altaffiliation[Now at ]{Department of Physics and Astronomy, University of Rochester, Rochester, New York 14627}} 

\newcommand{\LSU}{\altaffiliation[Now at ]{Department of Physics and Astronomy, Louisiana State University, Baton Rouge, LA 70803}} 

\author{F.~P.~An}\ZSU
\author{W.~D.~Bai}\ZSU
\author{A.~B.~Balantekin}\Wisconsin
\author{M.~Bishai}\BNL
\author{S.~Blyth}\NTU
\author{G.~F.~Cao}\IHEP
\author{J.~Cao}\IHEP
\author{J.~F.~Chang}\IHEP
\author{Y.~Chang}\NUU
\author{H.~S.~Chen}\IHEP
% ...
\collaboration{Daya Bay Collaboration}

\date{\today}

\begin{abstract}
We present a new determination of the smallest neutrino mixing angle $\theta_{13}$ and the mass-squared difference \dmtt\ using a final sample of \TOTALIBD\  inverse beta-decay (IBD) candidates with the final-state neutron captured on gadolinium.
This sample was selected from the complete data set obtained by the Daya Bay reactor neutrino experiment in 3158 days of operation.
Compared to the previous Daya Bay results, selection of IBD candidates has been optimized, energy calibration refined, and treatment of backgrounds further improved. 
The resulting oscillation parameters are $\mmthet =\SINTONETHREE$, $\mmdmtt = \DMTHREETWONORMAL$ for the normal mass ordering or $\mmdmtt = \DMTHREETWOINVERTED$ for the inverted mass ordering.
\end{abstract}


\maketitle

Neutrino oscillation has been firmly established by multiple observations since its discovery in 1998~\cite{Super-Kamiokande:1998kpq}. 
As this phenomenon is not required
by the Standard Model, it offers opportunities to search for new interactions and physical principles.
The three-neutrino paradigm of neutrino oscillation can be parametrized by three mixing angles, two mass-squared differences, and a $CP$ phase~\cite{ParticleDataGroup:2020ssz}.
This framework has been very successful in explaining most of the observations
made with accelerator, atmospheric, reactor and solar neutrinos. 
Our knowledge of the smallest neutrino mixing angle $\theta_{13}$ has been steadily improving since the first definitive determination in 2012~\cite{DayaBay:2012fng}. 
Besides being the best-measured neutrino mixing angle at present,  
precise knowledge of $\theta_{13}$ is important for testing the three-neutrino paradigm of neutrino mixing and as an invaluable input to model-building and to other experiments, most notably in 
resolving the neutrino mass hierarchy~\cite{JUNO:2021vlw} and the search for $CP$ violation in neutrino oscillation~{\cite{NOvA:2019cyt,T2K:2021xwb}.

Nuclear reactors produce low-energy electron antineutrinos, $\overline{\nu}_e$s, that are ideal for determining $\theta_{13}$ and the
mass-squared difference \dmtt\ through the study of $\overline{\nu}_e$ disappearance. 
This is best accomplished by comparing the energy spectra obtained with identically designed detectors positioned at different distances from the reactors. 
This relative approach cancels the uncertainties in the absolute detection efficiency that are correlated between detectors and heavily suppresses the effect of the uncertainty in the reactor $\bar{\nu}_e$ flux determination, thus enabling precision measurement of the oscillation parameters.
The $\overline{\nu}_e$s are detected via the inverse beta-decay reaction 
(IBD),  $\overline{\nu}_e + p \to e^+ + n$, with the 
kinetic-energy loss and annihilation of the positron giving rise to
a prompt-energy ($E_p$) signal, and the subsequent neutron capture to a delayed-energy ($E_d$) signal. 
The energy of the $\overline{\nu}_e$, $E_{\overline{\nu}}$, central to
measurements of neutrino oscillation,  is inferred 
from $E_p$ with $E_{\overline{\nu}} \approx E_p$ + 0.78~MeV.

In this Letter we report a new measurement of \thet\ and \dmtt\ using a final sample of \TOTALIBD\ IBD candidates with the final-state neutron captured on gadolinium (n-Gd) acquired by the Daya Bay reactor neutrino experiment in 3158 days of operation.

We utilized up to eight antineutrino detectors (ADs) to detect $\overline{\nu}_e$s emitted from three pairs of 2.9-GW$_{th}$ reactors at the Daya Bay-Ling Ao nuclear power facility in Shenzhen, China.  
The ADs were installed in three underground experimental halls, EH1, EH2, and EH3,  having a flux-averaged baseline of about 500~m, 500~m, and 1650~m from the reactors, respectively. 
To suppress ambient radiation, the ADs were submerged in water pools. Each pool was optically divided to function as inner (IWS) and outer (OWS) water Cherenkov detectors for detecting cosmic-ray muons. 
Four layers of Resistive Plate Chambers (RPCs) covering the top of each water pool 
provided another independent muon detector.
IBD events were detected with 20 tonnes of liquid scintillator doped with 0.1\% gadolinium  
by weight (GdLS) in each AD~\cite{Yeh:2007zz,DING2008238,BERIGUETE201482}.
The GdLS was contained in a 3-m-diameter acrylic cylinder enclosed inside a 4-m-diameter acrylic cylinder filled with 22 tonnes of undoped liquid scintillator (LS). 
Optical photons generated in the scintillator were detected with 192 photomultiplier tubes (PMTs) covering the barrel surface of the AD~\cite{DayaBay:2015meu}. 
The PMTs were arranged in 8 horizontal rings and 24 vertical columns.
Highly reflective disks sandwiching the 4-m acrylic vessel were used to enhance the detection efficiency of scintillation photons. 
Radioactive sources and LEDs were stored in three automatic calibration units (ACUs) on top of each AD~\cite{LIU201419,Liu:2015cra}. 
Detailed information of the experiment can be found in Refs.~\cite{DayaBay:2014cmr,DayaBay:2015kir}. 
For each AD, a cylindrical coordinate system with the vertical $z$-axis being the symmetry axis and $z$ = 0 at the AD center was used.

The Daya Bay experiment was operated with three different configurations of ADs in the three EHs. 
From 24 December 2011 to 28 July  2012 (217 days), the experiment ran in an initial six-AD configuration with 2 ADs in EH1, 1 AD in EH2 and 3 ADs in EH3 that resulted in the first observation of 
$\overline{\nu}_e$ disappearance at ${\cal O}(1\ {\rm km})$ baselines~\cite{DayaBay:2012fng}. 
An AD was added to both EH2 and EH3 during the summer of 2012 and this eight-AD configuration was operated from 19 October 2012 until 20 December 2016 (1524 days). 
Seven-AD operation occurred from 26 January 2017 until 12 December 2020 (1417 days) with one AD in EH1  re-purposed for liquid scintillator R\&D for the JUNO experiment~\cite{JUNO:2020bcl}.  

The results presented in this Letter are based on the data collected in the three configurations. 
Throughout the entire data analysis process, multiple groups within the collaboration provided validation and cross-checks. 

Details of the analysis process and techniques can be found in Refs.~\cite{DayaBay:2016ggj, DayaBay:2018yms}. 
In this Letter we focus on the improvements to the analysis techniques. 


Accurate and precise measurement of the prompt energy $E_p$ is essential for extracting the oscillation parameters from the spectra. 
After the gain of each PMT was calibrated with the single-photoelectron peak from dark noise, a correction for the non-linear response of the electronics was applied to each channel. 
This correction was derived from the waveform output from a 
flash-ADC readout system running in parallel with the default ADC system of EH1-AD1 in 2016~\cite{Huang:2017abb}.  
The observed charge profile was then used to reconstruct the position of the event using Reconstruction B in Ref.~\cite{DayaBay:2016ggj}. 

To obtain the reconstructed energy (E$_{rec}$), an additional correction to the non-uniform detector response was applied to account for a few non-functional 
PMTs toward the end of data collection. 
We used the energy deposited by spallation neutron capture on Gd in the GdLS and delayed $\alpha$-particles from correlated decays of natural radioactivity, ${}^{214}$Bi$\to$${}^{214}$Po$\to$${}^{210}$Pb, in the LS to determine this 
additional position-dependent correction. 
The active volume of each AD was divided into 100 voxels in $z$ and $r^2$, where $r$ is the radial distance from the $z$-axis.
For each voxel, the correction was defined as the ratio of the reconstructed energy to the reconstructed energy averaged over the entire GdLS volume. 
The temporal dependence of this correction was accommodated by two calibration periods, before and after 31 March 2017. 
The largest additional per-voxel correction was about 3\%.

In this study, the prompt energy was obtained by directly correcting 
E$_{rec}$ for the non-linear response of the LS 
which was determined from calibration~\cite{DayaBay:2019fje}.
Weekly calibration was performed by remotely lowering the calibration sources into the ADs from the ACUs.
Specialized calibration runs were taken during the re-configuration periods~\cite{Huang:2013uxa}. 
The positron response model of Ref.~\cite{DayaBay:2019fje} was updated, taking into account the 
measured responses of $\gamma$-rays from various sources and electrons from $\beta$-decay of cosmogenic $^{12}$B of the full dataset as inputs. 
The best-fit model had a Birks' coefficient $k_B$ = 0.0143~g/cm$^2$/MeV for the quenching effect and $k_C=0.023$ for the contribution of Cherenkov radiation to the non-linearity; both parameters agreed well with the previous result~\cite{DayaBay:2019fje}. The improved energy response model for the positron achieved a precision of $< 0.5$\% for $E_p >$ 2 MeV.

IBD candidates were selected with the following criteria.
Events caused by spontaneous light emission of the PMTs, so-called flashers, were removed.
Candidates must have a prompt-like signal with 0.7 MeV $< E_p <$ 12 MeV 
separated by 1 to 200 $\mu$s from a delayed-like signal with 6 MeV $< E_d <$ 12 MeV. 
Candidate pairs were vetoed if their delayed-like events occurred (i) within a (-202 $\mu$s, 600 $\mu$s) time-window with respect to an IWS or OWS trigger with a PMT-hit multiplicity (nHit) $>12$, or (ii) within a (-202 $\mu$s, 410 $\mu$s) time-window with respect to an IWS trigger with 6 $<$ nHit $\le$ 12, or 
(iii) within a (-202 $\mu$s, 1400 $\mu$s) time-window with respect to triggers in the same AD with energy between 20 MeV and 2~GeV or (iv) within a (-202 $\mu$s, 0.4 s) time-window with respect 
to triggers in the same AD with energy higher than 2 GeV. 
This targeted muon veto efficiently removed spurious triggers that followed a muon as well as most muon-induced spallation products and muon decays. 
To remove any ambiguity in the 
candidate-pair selection, no additional AD triggers with energy between 0.7 MeV and 20 MeV were allowed within (-400 $\mu$s, 200 $\mu$s) of the delayed candidate.

A new source of flashers was observed in the 7-AD operation period that were not suppressed by the previous criteria~\cite{DayaBay:2016ggj}. 
Additional selection criteria targeting the characteristic 
charge pattern and temporal distribution of these new flashers were devised that rejected over 99\% of this instrumental background with an IBD selection efficiency over 99.99\%. 

The selected IBD candidates consisted of genuine IBD and background events. 
The background comprised uncorrelated accidental pairs, and correlated prompt-and-delayed signals coming from fast neutrons, $\beta$-n decays of 
spallation ${\rm ^{9}Li/^{8}He}$, neutrons leaking from the $^{241}$Am-$^{13}$C calibration sources and ${\rm ^{13}C(\alpha,n)^{16}O}$ with the $\alpha$ coming from natural radioactivity. 
The latter two correlated backgrounds and the accidental background, detailed in Ref.~\cite{DayaBay:2016ggj}, did not require any improved treatment in this analysis. 
The muon detection efficiency of the IWS and OWS dropped with time due to the gradual loss of functional PMTs near the top of the water pools, particularly in the 7-AD period. 
With this loss of detection efficiency, a new background, dubbed ``muon-x'' (described below), became apparent.

The largest correlated background is $\beta$-n decay of cosmogenic radio-isotopes  $^9$Li and $^8$He. 
To determine this background, muons were paired with all IBD candidates within 
$\pm$2 seconds. 
To improve discrimination of ${}^{9}{\rm Li}/{}^8{\rm He}$ from other processes, candidate events were separated into several samples based on the visible energy 
deposited by the muon ($E_\mu$) in the AD and the distance between the 
prompt and delayed signals, $\Delta r$. 
The rates and energy spectra of the dominant cosmogenic radio-isotopes were extracted with a simultaneous fit to 12 two-dimensional histograms defined by the different muon samples in the three experimental halls for the two $\Delta r$ regions with a 
probability density function $\phi(E_p,\Delta t)$,
where $\Delta t$ is the time difference of the prompt-energy signal 
and the muon~\cite{supplemental}. 
The distribution in $\Delta t$ is described by a sum over radio-isotopes, taking into account the known isotope lifetimes, and a term for uncorrelated muon-IBD pairs that is well-constrained by the pairs in the region $\Delta t<0$ s. 
Since the lifetimes of $^9{\rm Li}$ and $^8{\rm He}$ are comparable, we simply measured the sum of these two radio-isotopes.
The $E_p$ distributions of the radio-isotopes were determined from the fit.
This method provides higher statistics and a better determination of the low-energy part of the $\beta$ spectrum of $^9{\rm Li}$/$^8{\rm He}$ than the previous determination~\cite{DayaBay:2016ggj} while reducing the rate uncertainty to less than 25\%. 

We determined the combined contribution of the fast-neutron and muon-x processes to the background.

Energetic neutrons generated by cosmic-ray muon interactions in the vicinity of the water pool can enter the active volume of an AD. 
Proton recoil from neutron scattering in the LS and the subsequent neutron capture on gadolinium constituted the prompt and delayed signals of this fast-neutron background and dominates the correlated events with $E_p >$ 12 MeV.
The energy spectrum was determined from prompt signals in coincidence with muons detected only by the OWS or RPC within $0.5\ \mu{\rm s}$ of a prompt candidate with 0.7 MeV $ < E_p < $ 250 MeV. 
The measured spectra were similar in shape among the three experimental halls and stable throughout all three operational periods.

The muon-x background was caused by low-$E_\mu$ muons 
that passed through the IWS undetected. These events typically consisted of the muon as the prompt signal and a Michel electron from muon decay or a spallation neutron as the delayed signal. 
Due to the decrease in efficiency of the IWS that occurred mainly during the 7-AD period, this background could not be removed with the IWS nHit $>\!12$ criterion, but was efficiently ($>$ 80\%) removed by vetoing events with a delayed signal less than 410 $\mu$s after a muon identified with an IWS nHit satisfying 6 $<$ nHit $\le$ 12, which led to a loss in livetime of less than 0.1\%.
The muon-x prompt- and delayed-energy spectra of the remaining muon-x background were approximated by the muon-x sample obtained with IWS ${\rm nHit} = 7$.

To determine the rate\sout{s} of these two backgrounds, the prompt-energy spectrum above 12 MeV and the delayed-energy spectrum of the 
IBD-candidate sample with the prompt energy extended to 
250 MeV were 
fitted to the spectra of the previously described OWS-tagged fast-neutron sample and the muon-x sample with IWS ${\rm nHit}=7$.
Their rates in the range of 0.7 MeV $ < E_p < $ 12 MeV were estimated by extrapolation. 
We found no muon-x background in the 6-AD period. In the 8-AD period, the rate of muon-x background was about 0.038 of that of the fast neutron in EH1, 0.0055 in EH2, and 0.023 in EH3. 
The ratios increased to 0.26, 0.15, and 0.53, respectively, in the 7-AD period, consistent with the reduction in the number of functional PMTs in the three IWSs.
The combined systematic uncertainty of these two backgrounds was estimated to be about 20\%,  
which is dominated by the fast neutron background. 
As validation,  the muon-x background was deduced by comparing the prompt- and delayed-energy spectra of data samples before the 7-AD period with and without masking the PMTs that failed subsequently.
Consistent results were obtained.

Table~\ref{tab:ibd} summarizes the IBD candidates and backgrounds for the final n-Gd sample. 
We obtained a total of 4.8 million IBD candidates at the near halls and 0.76 million at the far hall with less than 2\% background.

\begin{table*}[!htb]
\caption{
Summary of IBD signal and background. 
Rates are corrected for the muon veto and multiplicity selection efficiencies $\varepsilon_{\mu}\times\varepsilon_{m}$. 
The sum of the fast neutron and muon-x background rates is reported as ``Fast n + muon-x". 
The AD numbering scheme reflects the time order of AD fabrication and deployment.
\label{tab:ibd}
}
  \begin{ruledtabular}{\footnotesize
\begin{tabular}{c cc@{\hskip 2pt} cc@{\hskip 2pt} cccc} 
  & \multicolumn{2}{c}{EH1}&\multicolumn{2}{c}{EH2}&\multicolumn{4}{c}{EH3} \\ 
\cline{2-3}\cline{4-5}\cline{6-9} 
  & AD1  & AD2  & AD3 & AD8 & AD4 & AD5 & AD6 & AD7 \\
\hline
\nuebar candidates & 794335	& 1442475 & 1328301 & 1216593 & 194949 & 195369	& 193334 & 180762 \\
DAQ live time [days] & 1535.111 & 2686.110 & 2689.880 & 2502.816 & 2689.156 & 2689.156 & 2689.156 & 2501.531  \\
$\varepsilon_{\mu}\times \varepsilon_{m}$ & 0.7743 & 0.7716 & 0.8127 & 0.8105 & 0.9513  & 0.9514 & 0.9512 & 0.9513\\
Accidentals [day$^{-1}$] & $7.11\pm0.01$ & $6.76\pm0.01$ & $5.00\pm0.00$ & $4.85\pm0.01$ & $0.80\pm0.00$ & $0.77\pm0.00$ & $0.79\pm0.00$ & $0.66\pm0.00$ \\
Fast n + muon-x [day$^{-1}$] & $0.83\pm0.17$ & $0.96\pm0.19$ & $0.56\pm0.11$ & $0.56\pm0.11$ & $0.05\pm0.01$ & $0.05\pm0.01$ & $0.05\pm0.01$ & $0.05\pm0.01$  \\
$^9$Li/$^8$He [AD$^{-1}$ day$^{-1}$] & \multicolumn{2}{c}{$2.92\pm0.78$} & \multicolumn{2}{c}{$2.45\pm0.57$} & \multicolumn{4}{c}{$0.26\pm0.04$} \\
$^{241}$Am-$^{13}$C [day$^{-1}$] & $0.16\pm0.07$ & $0.13\pm0.06$ & $0.12\pm0.05$ & $0.11\pm0.05$ & $0.04\pm0.02$ & $0.04\pm0.02$ & $0.04\pm0.02$ & $0.03\pm0.01$ \\
$^{13}$C($\alpha$, n)$^{16}$O [day$^{-1}$] & $0.08\pm0.04$ & $0.06\pm0.03$ & $0.04\pm0.02$ & $0.06\pm0.03$ & $0.04\pm0.02$ & $0.04\pm0.02$ & $0.03\pm0.02$ & $0.04\pm0.02$ \\ \hline
\nuebar rate [day$^{-1}$] & $657.16\pm1.10$ & $685.13\pm1.00$ & $599.47\pm0.78$	& $591.71\pm0.79$ & $75.02\pm0.18$ & $75.21\pm0.18$ & $74.41\pm0.18$	& $74.93\pm0.18$ \\
\end{tabular}
  }\end{ruledtabular}
\end{table*}

The $\overline{\nu}_e$ flux without oscillation at each AD was predicted by 
using the thermal-power data and fission fractions of each fuel cycle, provided by the power plant operator, as a function of burn-up.
The power data carried an uncorrelated uncertainty of 0.5\% per core, while a 0.6\% uncorrelated uncertainty per core in the $\overline{\nu}_e$ yield was introduced by the uncertainties of the fission fractions. 
Due to the nature of the near-far relative measurement, 95\% of the uncorrelated uncertainty of each core cancelled and extraction of the oscillation parameters was insensitive to the spectral shape of the
no-oscillation prediction.

The detector-related uncertainties have been presented in Ref.~\cite{DayaBay:2016ggj}.
Detection efficiency uncertainties that are correlated
between detectors did not contribute to this near-far relative measurement.
The total uncorrelated uncertainty in the detection efficiency remained at 0.13\%.
The largest contribution of 0.10\% coming from the fraction of neutrons captured on gadolinium was obtained by comparing the capture-time distributions of the ADs. 
The next largest uncorrelated uncertainty of 0.08\% in the delayed-energy 
selection criterion was due to a 0.2\% spread in the relative energy scale among the ADs. 
The relative detection efficiency estimate was validated by comparing the $\overline{\nu}_e$ rates of neighboring ADs in each EH for each data-taking period~\cite{DayaBay:2012aa}.  
The rates were consistent with the predictions that took the tiny variations 
in the baseline and number of protons into account.
Furthermore, no significant deviation in the spectral distributions among the ADs in the same experimental hall was found.

We extracted the oscillation parameters using
the survival probability of three-flavor oscillation given by
\begin{eqnarray}
P & = & 1 - {\rm cos}^4\theta_{13}{\rm sin} ^22\theta_{12}{\rm sin^2}\Delta_{21} \nonumber \\
&&- \rm{sin}^22\theta_{13}\left({\rm cos}^2\theta_{12}{\rm sin}^2\Delta_{31} + {\rm sin}^2\theta_{12}{\rm sin}^2\Delta_{32}\right)
\end{eqnarray}\label{eq:survive}
\noindent where $\Delta_{ij} = 1.267\Delta m^2_{ij} L/E$ with $\Delta m^2_{ij}$ in eV$^2$, $L$ is the baseline in meters between an AD and a reactor core and $E$
is the energy of the $\overline{\nu}_e$ in MeV.
We used ${\rm sin} ^2\theta_{12} = 0.307 \pm 0.013$ and $\Delta m^2_{21} = (7.53 \pm 0.18) \times 10^{-5}~{\rm  eV}^2$~\cite{ParticleDataGroup:2020ssz}.
Alternatively, for short baselines of a few kilometers, the survival probability can be parametrized as
\begin{eqnarray}
P & = & 1 - {\rm cos}^4\theta_{13}{\rm sin}^22\theta_{12}{\rm sin^2}\Delta_{21} - \rm{sin}^22\theta_{13}{\rm sin}^2\Delta_{ee}.
\label{eq:m2ee}
\end{eqnarray}
\noindent Here, the effective mass-squared difference \dmee\  is
related to the wavelength of the oscillation observed at Daya Bay, 
and is independent of the choice of neutrino mass ordering as well as the value and uncertainty of the mixing angle $\theta_{12}$~\cite{DayaBay:2016ggj}. 

We adopted fitting Method B reported in Ref.~\cite{DayaBay:2016ggj} to extract the oscillation parameters. 
The fit minimized a $\chi^2$ function defined as~\cite{supplemental}:
\begin{equation}
\chi^2(\theta_{13}, \Delta m^2, {\bm \nu}) = 
\chi^2_{\rm stat}(\theta_{13}, \Delta m^2, {\bm \nu}) + 
\chi^2_{\rm syst}({\bm \nu})
\end{equation}
\noindent where $\chi^2_{\rm stat}$ is the standard statistical term that compares all the measured background-subtracted prompt-energy spectra with the predictions. 
For each period of operation, the spectrum of each AD was divided into 26 bins.
The predictions were derived from the calculated reactor $\overline{\nu}_e$ flux, survival probability, IBD cross section~\cite{Vogel:1999zy} and detector response obtained with a detailed Geant4-based simulation~\cite{AGOSTINELLI2003250,ALLISON41610988,ALLISON2016186}.
The term $\chi^2_{\rm syst}({\bm \nu})$ contains the 
detector and background systematic uncertainties as pulls of the nuisance parameters expressed as a vector ${\bm \nu}$.

Figure~\ref{fig:contours} shows the covariance contours in the \dmee-\thet\  space.
The best-fit point with $\chi^2$/ndf = 559/517 yields $\mmthet =\SINTONETHREE$, and $\mmdmtt = \DMTHREETWONORMAL$ for the normal mass hierarchy or $\mmdmtt = \DMTHREETWOINVERTED$ for the inverted mass hierarchy.
Using Eq.~\ref{eq:m2ee}, we obtained $\mmthet = 0.0852 \pm 0.0024$ and $\mmdmee = \DMEE$ with the same reduced-$\chi^2$ value.
Results determined with the other fitting methods described in Ref.~\cite{DayaBay:2016ggj} were consistent to $<\!0.2$ standard deviations.

\begin{figure}[htb]
\includegraphics[width=\columnwidth]{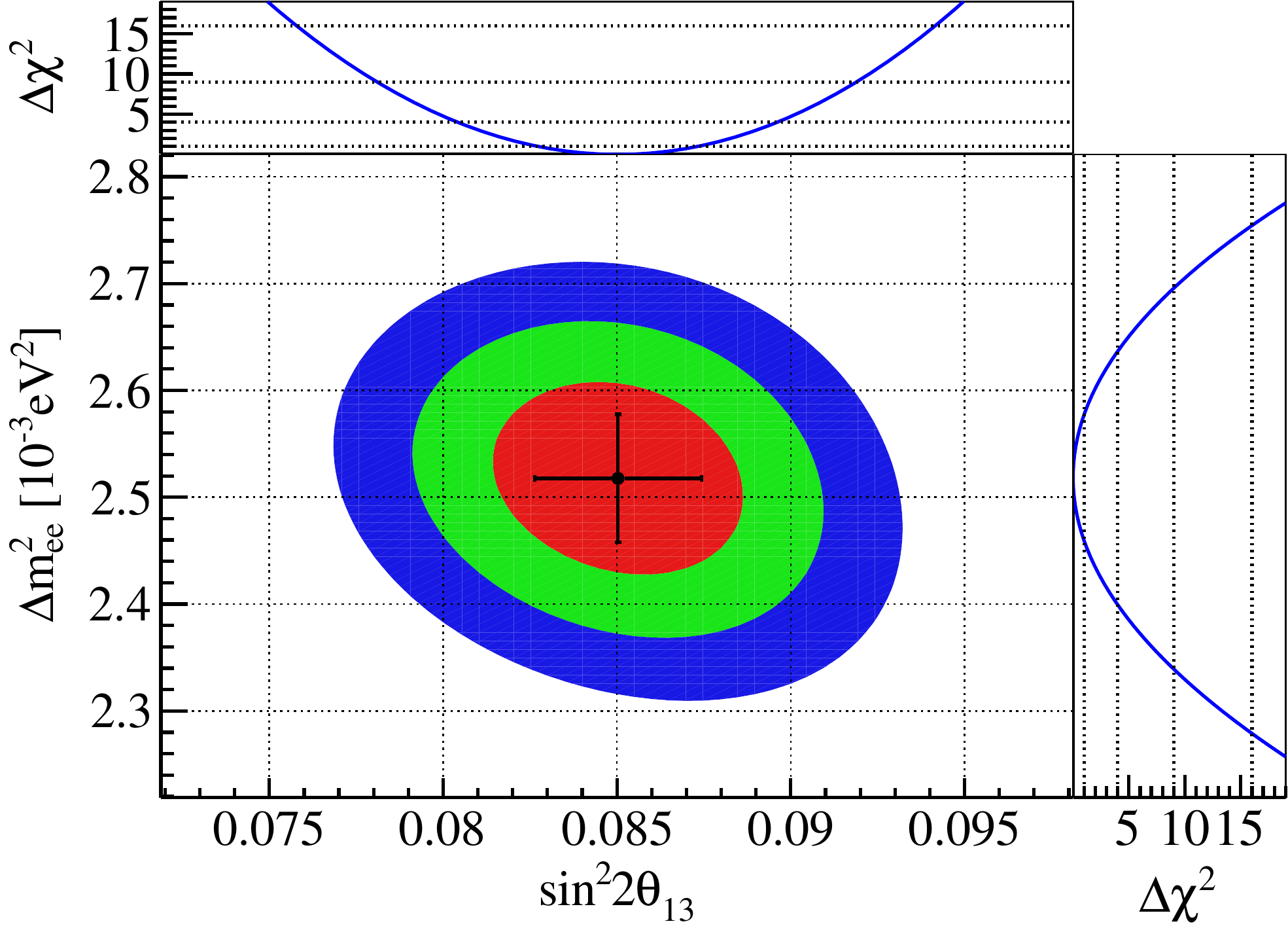}
\caption{\label{fig:contours} Error ellipses in the 
\dmee-\thet\ space with the best-fit point indicated. 
The error bars display the one-dimensional one-standard deviation confidence intervals. 
The colored contours correspond to one, two, and three standard deviations.
The $\Delta \chi^2$ distributions are also shown. 
These one-dimensional distributions were obtained by determining the smallest 
$\Delta \chi^2$ value after scanning through
\dmee\ (\thet\ ) for a given \thet\ (\dmee\ ). }
\end{figure}

The best-fit prompt-energy distribution is in excellent agreement 
with the observed spectra in each experimental hall, as shown in Fig.~\ref{fig:spectra}. 

\begin{figure*}[htb] 
\includegraphics[width=0.68\columnwidth]{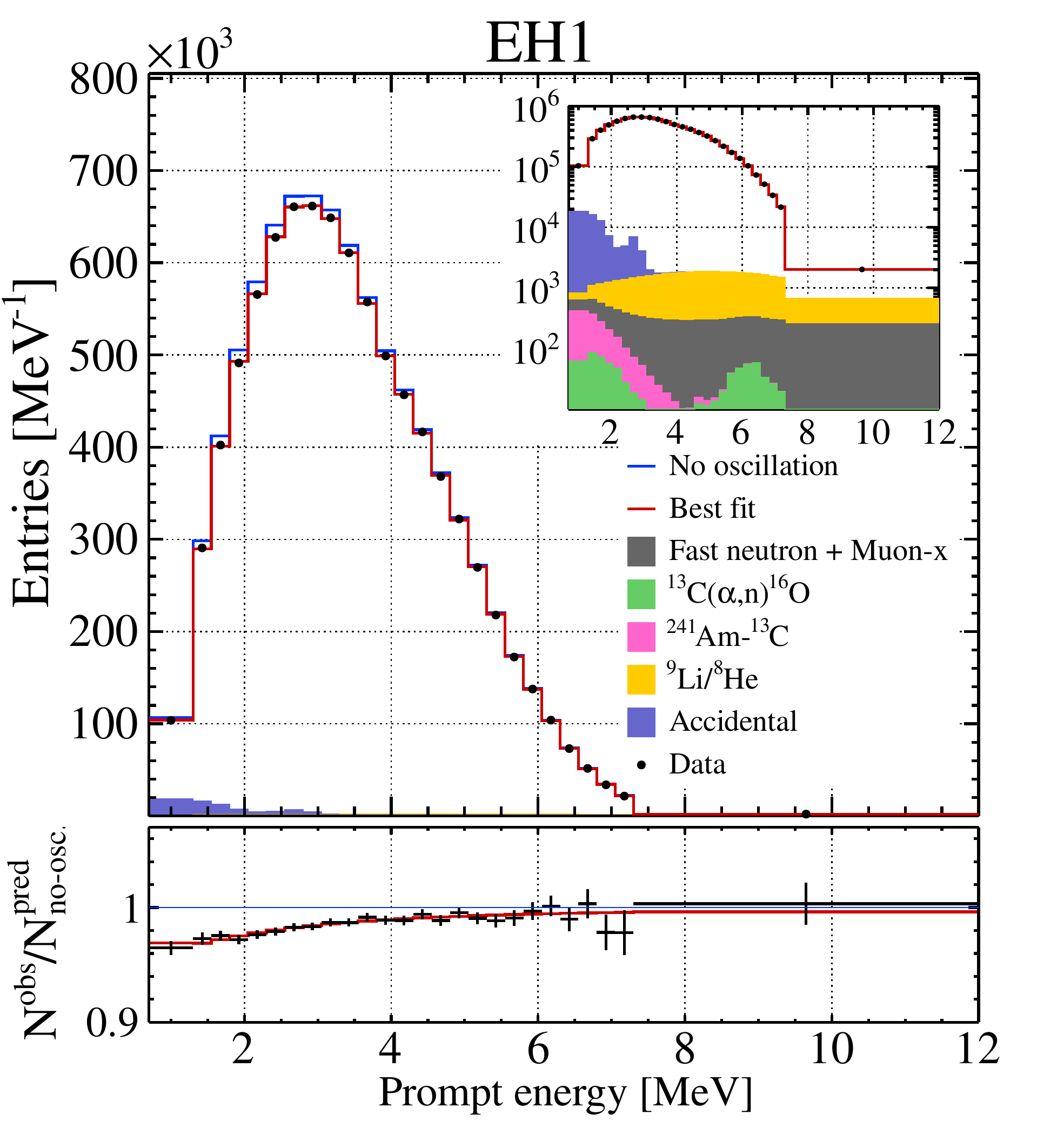}
\includegraphics[width=0.68\columnwidth]{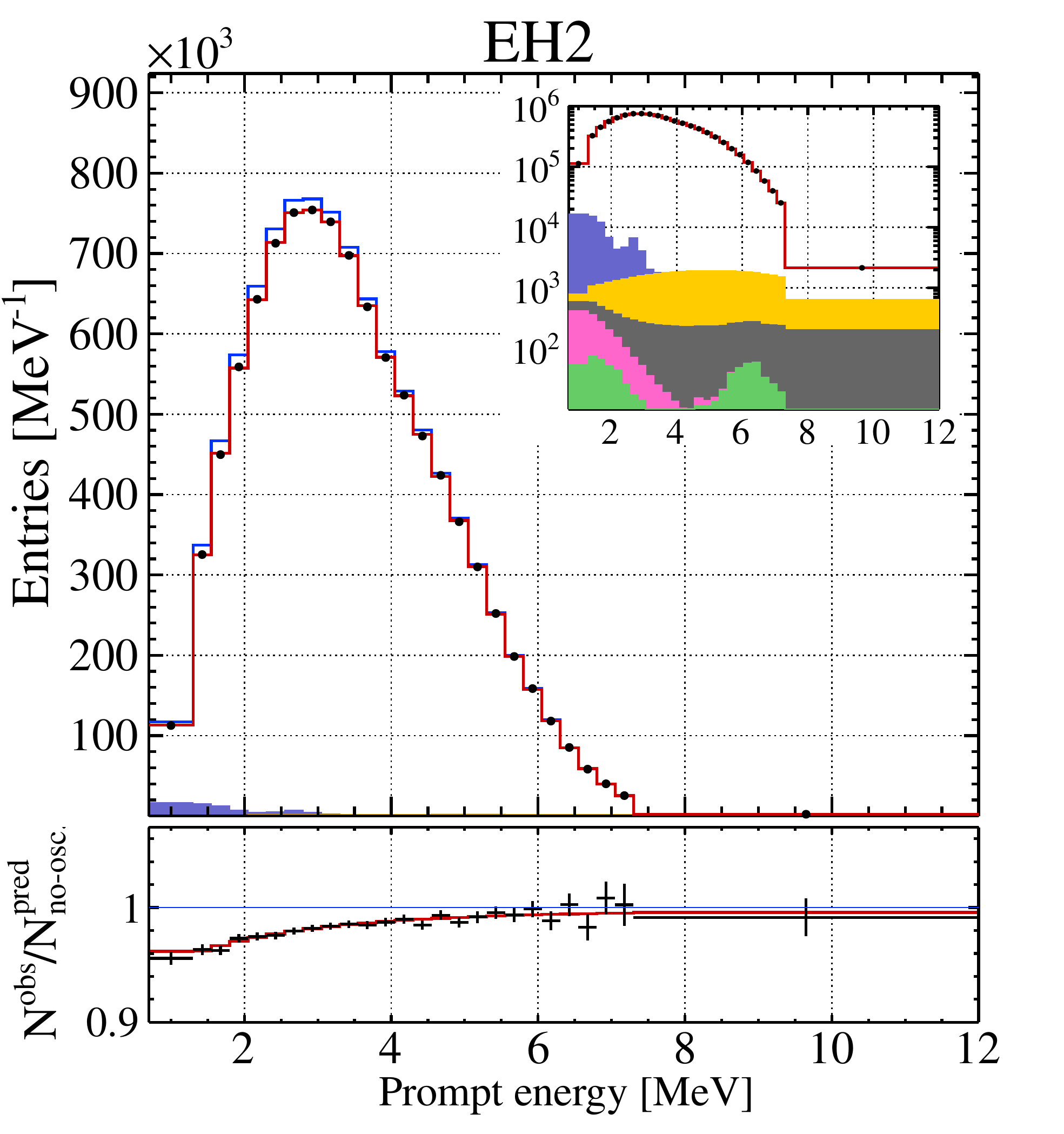}
\includegraphics[width=0.68\columnwidth]{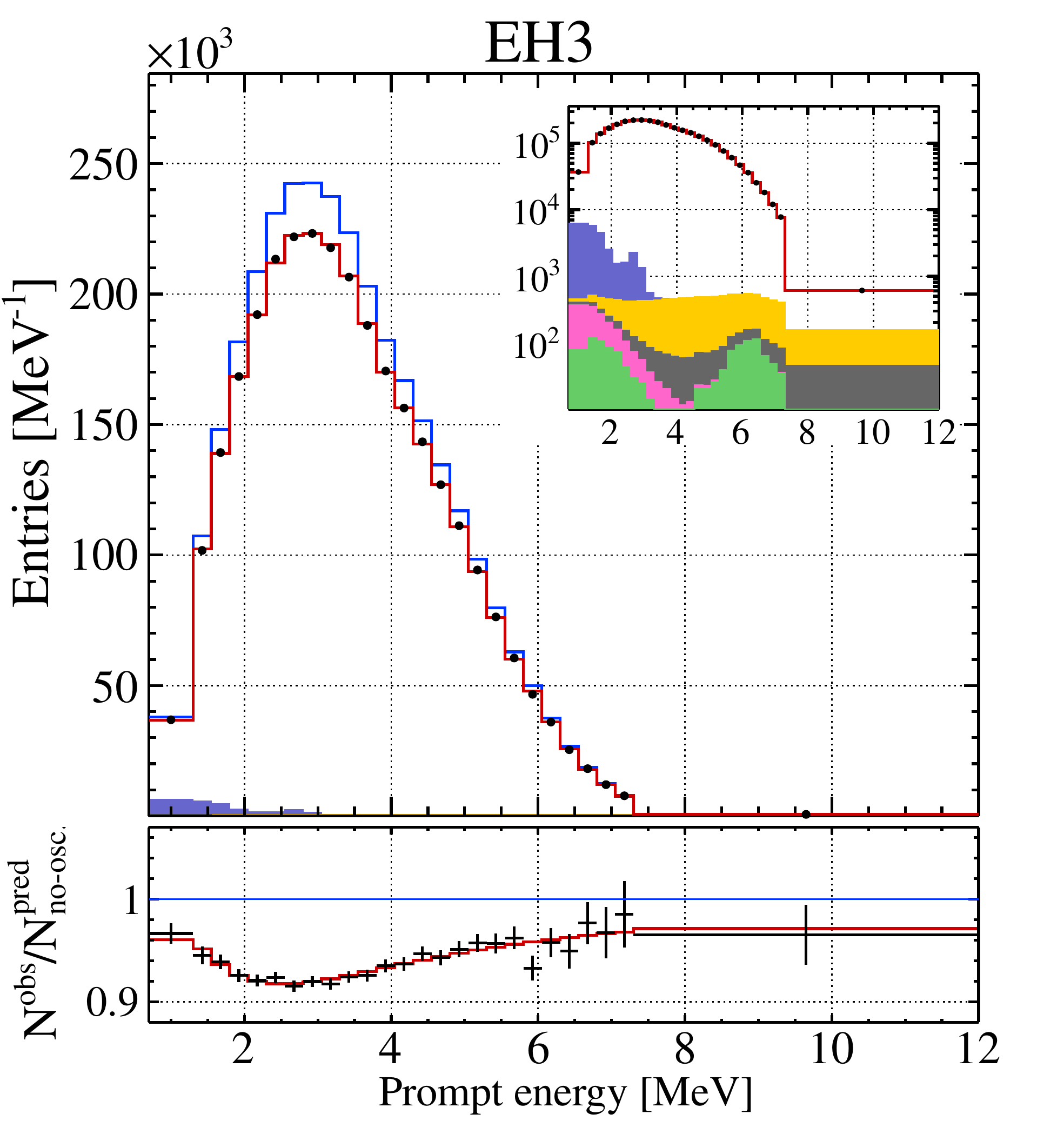}
\caption{\label{fig:spectra} The measured prompt-energy spectra of EH1, EH2 and EH3 with the best-fit and no-oscillation curves superimposed in the upper panels. 
The shape of the backgrounds are apparent in the spectra with a logarithmic ordinate shown in the insets. 
The lower panels shows the ratio of the observed spectrum to the predicted no-oscillations distribution. The error bars are
statistical.
}
\end{figure*}
Figure~\ref{fig:LoverE} depicts the normalized signal rate of the three halls as a function of $L_{eff}/\langle E_{\overline{\nu}_e}\rangle$ with the best-fit curve superimposed, where $L_{eff}$ and ${\langle  E_{\overline{\nu}_e}\rangle}$ are the effective baseline and average $\overline{\nu}_e$ energy, respectively~\cite{DayaBay:2016ggj}. 
The oscillation pattern related to $\theta_{13}$ is unambiguous.
\begin{figure}[htb]
\includegraphics[width=\columnwidth]{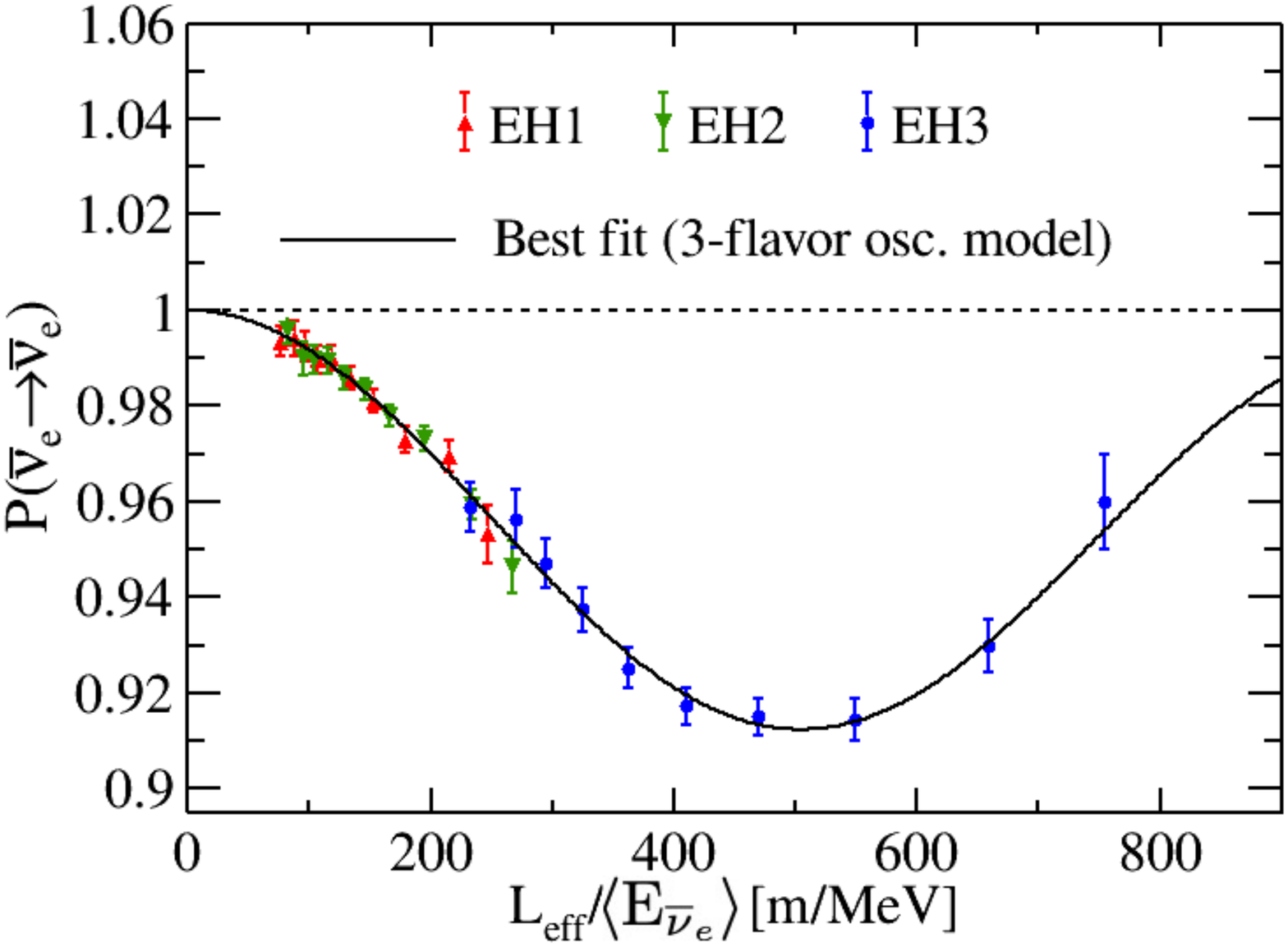}
\caption{\label{fig:LoverE} 
Measured disappearance probability as a function of the ratio of the effective baseline $L_{eff}$ to the mean antineutrino energy $\langle E_{\overline{\nu}_e}\rangle$.
}
\end{figure}

The present improved result in \thet\  is consistent with 
our previous determinations~\cite{DayaBay:2012fng, DayaBay:2016ggj, DayaBay:2018yms} and agrees with other measurements of reactor $\overline{\nu}_e$ disappearance by RENO~\cite{RENO:2018dro} and Double Chooz~\cite{DoubleChooz:2019qbj,DoubleChooz:2020vtr} 
as well as electron neutrino and antineutrino appearance measurements by T2K~\cite{T2K:2021xwb}.
Daya Bay's measured \dmtt\  is consistent with the results of NOvA~\cite{NOvA:2019cyt}, T2K~\cite{T2K:2021xwb}, 
MINOS/MINOS+~\cite{MINOS:2020llm}, IceCube~\cite{IceCube:2017lak} and SuperK~\cite{Super-Kamiokande:2017yvm} that were obtained with muon 
(anti)neutrino disappearance. 
The agreement in \thet\  and \dmtt\  between Daya Bay measurements using $\overline{\nu}_e$ and the muon neutrino and antineutrino determinations provides strong support of the three-neutrino paradigm. 

To conclude, we have presented a new determination of \thet\  with a precision of 2.8\% and the mass-squared differences reaching a precision of about 2.4\%. 
The reported \thet\ will likely remain the most precise measurement of $\theta_{13}$ in the foreseeable future and be crucial to the investigation of the mass hierarchy and $CP$ violation in neutrino oscillation.

The Daya Bay experiment is supported in part by the Ministry of Science and Technology of China, the U.S. Department of Energy, the Chinese Academy of Sciences, the CAS Center for Excellence in Particle Physics, the National
Natural Science Foundation of China, the Guangdong provincial government, the Shenzhen municipal government, the China General Nuclear Power Group, the Research Grants Council of the Hong Kong Special Administrative Region
of China, the Ministry of Education in Taiwan, the U.S. National Science Foundation, the Ministry of Education, Youth, and Sports of the Czech Republic, the Charles University Research Centre UNCE, and the Joint Institute of
Nuclear Research in Dubna, Russia. 
We acknowledge Yellow River Engineering Consulting Co., Ltd., and China Railway 15th Bureau Group Co., Ltd., 
for building the underground laboratory. We are grateful for the cooperation 
from the China Guangdong Nuclear Power Group and China Light \& Power Company.


\bibliographystyle{apsrev4-1}
\bibliography{OtherCitations,DayaBayCitations,DoubleChoozCitations,RENOCitations,supplemental}

\end{document}